Analytical methods of asymmetry double sine-Gordon equation in infinite one-dimensional system
Nan-Hong Kuo
C. D. Hu
Departement of physics, National Taiwan University, Taipei, Taiwan, R.O.C.

Abstract
Traditionally, Double Sine-Gordon Equation (DSGE) is seen as a nonintegrable equation. That means we cannot find general solutions in asymmetry DSGE. In this paper, we develop analytical method to solve this equation by Möbius transformation. And finally, this can reduce the problem to find roots of polynomial of four degree in one element. We have known this can be solved by square formally because its degree less than five. Although complexity as a solution, but in this sense, we can say we formally solve this nonintegrable equation.

## 1. Introduction:

As for our pervious paper [1], [17], the more accutate equation is Double Sine-Gordon Equation (DSGE) to solve one-dimensional spin chain problem. Solving asymmetry DSGE can help us understand how solution evolve with adiabatical parameter and so we can understand its original physical quantities changed with adiabatical parameter. About our main interest original of double sine-grodon equation-quantum spin pump, we can see [1], [2], [3], [17]. The mathematial solution of this equation, we can see , for example, [4], [5], [6]. We also can see [7], [8], [9], [10] to see how to do bosonization to calaulate one-dimensional quantum physics quantities. Bosonization is our main tool to transform our interrst one-dimensional antiferromagnetism quantum spin chain to asymmetry double sine-gordon equation.

Alough in [1] and [17] , we do Sine-Gordon Equation (SGE) due to good approxination of renormation group analysis in one dimension, but in later part about numerical calaulate of [2], the author still use original physical system, which means he use DSGE instead SGE and have the evident of spin pump due to edge state. And in reality physical system, even irrelevent operator cannot ignore completely because we always tangle finite system. In this paper, we build more accute model, DSGE in Infinite system, than [1].

SGE as one of the famous integrable equation, not only we have known all its solutions, but also known the relations between them. See our recently work, [18] and [19]. We have modular and Landen transformation between these solutions. DSGE as a nonintegrable equation, but a closely equation of SGE,  does it exist similar transfornation? Fortually, it does by Möbius transformation. And finally, we will give the analytical method to solve them. In this sense, DSGE seems not a nonintegrable equation at all.

But firstly, we also concern what other people study this problem ( asymmetry DSGE)? I mainly [12],[13],[14],[15]. Because DSGE is not an integrable equation traditionally, so the works of them are using perturbation method -Form factor perturbation theory (FFPT) to sovle asymmetry DSGE using intergrable equation as basis, like what we study in perturbation chapter in Quantum Mechanics books. And here integrable model means SGE. We have well known its spectrum. Alough this idea is good, but we must remember, for example, see [10], [16], that perturbation is false in one dimension because their excitation is not quasi-particle like in higher dimension, but the collective excitation in one dimension. This phenomena can be describely by susceptibility, which measure the response due to external parameter. Another reason is the well known: Renormalization group in SGE.

But the perturbation in one dimension is false in state, not for energy. That is why [12]~[15] only list energy correction. we also can see §1.2 of [10], it also list energy correction, like sense of FFPT, but more semiclassical sense. Due to the reason above, if we want to get the state of asymmetry DSGE, we actually don't use perturbation theory, especially for infinite length in one dimension. We must "exactly' solve this nonintegrable equation.

$$\theta_{tt} - \theta_{xx} + \sin(\theta + \phi) + 2\eta \sin 2\theta = 0 \qquad (1)$$

In Section **2**, I list classical solutions of DSGE depend on the second potential coefficient $\eta$.

It is helpful for us to further study asymmetry DSGE. In Section **3** is our main contribution. We will give the analytical method using Möbius transform. In Section **4**, We give discussion and results. In Appendix, I introduce Möbius transformation which is related to my technique to asymmetry DSGE.

## 2. Classical solutions of Double Sine-Gordon Equation without phase $\varphi$:

In this section, we briefly review the solutions, potential and energy of the DSGE with $\phi = 0$ in order to establish the notations and set the stage for developing our method. We shall use the notations of [4] and also point out some errors in that work. As mentioned in [4], the Hamiltonian of the DSGE can be viewed as a chain of physical pendula joined by torsion springs

$$H = \int dx \{ \frac{p_\theta^2}{2I} + \frac{\Gamma}{2}\theta_x^2 - V_0[\cos(\theta + \phi) + \eta\cos 2\theta]\} \tag{2}$$

where $\theta$ is the angular coordinate, $p_\theta$ is the conjugate momentum, $I$ is the moment of inertia, $\Gamma$ is the torsional constant, and $V_0$ is the external potential. The DSGE we are interested in is (1). It is the equation of motion of the Hamiltonian

$$H = \int_{-\infty}^{\infty} [\frac{\theta_t^2}{2} + \frac{\theta_x^2}{2} - \cos(\theta + \phi) - \eta\cos 2\theta]dx \tag{3}$$

which can be obtained with rescalings of the time-space coordinates. In order to calculate the total energy, one can Integrate both sides of eq. (1) and get

$$\frac{1}{2}(\frac{d\theta}{ds})^2 + \cos(\theta + \varphi) + \eta\cos(2\theta) = S \tag{4}$$

where $s = x - t$. The negative sum of the last two terms can be viewed as the potential

$$V(\theta; \eta, \phi) = -\cos(\theta + \varphi) - \eta\cos(2\theta) \tag{5}$$

and so $S$ is called the "action". We can show that

$$S = -V_{\min} \tag{6}$$

The solutions of kinks and bubbles approach constant when $s \to \pm\infty$. Hence, $d\theta/ds|_{s=\pm\infty} = 0$ and $S = -V(\theta(s \to \pm\infty))$. Since $(d\theta/ds)^2/2 = V(\theta) - V(\theta(s \to \pm\infty)) \geq 0$ for any $s$, we conclude that $S = -V(\theta(s \to \pm\infty)) = -V_{\min}$. More specifically, whether the minimum is an absolute minimum or a relative minimum depends on what type of the solution is.

The energy for any value of $\phi$ can be calculated quite simply if forms of the solutions are known. We substitute $s$ with $x$ as we study the static kinks. From (4), the calculation of energy can be performed with the a method similar to that of [2]:

$$\frac{d\theta}{dx} = \sqrt{2[S - \cos(\theta + \varphi) - \eta\cos 2\theta]} \tag{7}$$

We define

$$V_s \equiv \int_{-\infty}^{\infty} \frac{\theta_x^2}{2} dx = \frac{1}{\sqrt{2}} \int_0^{2\pi} \sqrt{S - \cos(\theta + \varphi) - \eta\cos 2\theta}\, d\theta \tag{8}$$

and

$$V_p \equiv \int_{-\infty}^{\infty} [-\cos(\theta + \varphi) - \eta\cos 2\theta]dx \tag{9}$$

so that $H = V_s + V_p$. From (4) we found that

$$V_p = -S \cdot x|_{-L}^{L} + V_s \tag{10}$$

But the first term on the right hand side of (10) diverges if we consider an infinite system. Thus we substract this trivial infinity from the Hamiltonian in (3). This is equivalent to shift the system to a new energy zero. As a result, in view of (3) in the static case,

$$H = 2V_s \tag{11}$$

An analysis of the potential term is in order. It is helpful in understanding how state evolves with respect to the phase $\varphi$. The potential of double sine-Gordon equation has the symmetry

$$V(\theta; \eta, \phi = \tfrac{\pi}{2}) = V(\tilde{\theta}; -\eta, \phi = 0) \tag{12}$$

where $\tilde{\theta} = \theta + \pi/2$. This implies that the solution of $\phi = \pi/2$ has the same form as that of $\phi = 0$ and $-\eta$. However, interestingly, the solutions can have completely different forms for $\phi = 0$ or $\phi = \pi/2$ when $|\eta| > \tfrac{1}{4}$ (see below).

In view of (5), we have

$$\begin{aligned}\frac{dV}{d\theta} &= \sin(\theta + \phi) + 2\eta \sin(2\theta) \\ &= \sin\theta(\cos\phi + \cot\theta\sin\phi + 4\eta\sin\theta) \\ &= \cos\theta(\sin\phi + \tan\theta\cos\phi + 4\eta\sin\theta)\end{aligned} \tag{13}$$

The point $\theta = \cos^{-1}(-1/4\eta)$ is one of the absolute maxima only if $\phi = 0$ and $\eta < -1/4$, and $\theta = \sin^{-1}(-1/4\eta)$ is one of the absolute minima only if $\phi = \pi/2$ and $\eta > 1/4$. Moreover, for $\eta < -1/4$ the relative minima are at $\theta = \pi/2 + 2n\pi$ and for $\eta > 1/4$ the relative maxima are at $\theta = 2n\pi$. The shape of the potential has critical influence on the form of the solution.

Specifially, when $\varphi = 0$, the analytical forms of solution and energy can be obtained. So can those forms at $\phi = \pi/2$, by applying the symmetry proporty in eq. (11). In the following we summerizethem in order to set up our analysis.

**Case 1.** $\eta < -\tfrac{1}{4}$

In this case the minima are at $\theta_{abs.\min} = \phi_0 + 2n\pi$ where

$$\phi_0 \equiv \arccos(\tfrac{-1}{4\eta}) \tag{14}$$

From above, we have $S = -V(\theta_{\min}) = -\eta - 1/8\eta$. The absolute maximum of $V(\theta)$ are located at $\theta_{abs.\max} = (2n+1)\pi$ while the relative maximum are located at $\theta_{rel.\max} = 2n\pi$ with $V(\theta_{abs.\max}) = 1 - \eta$ and $V(\theta_{rel.\max}) = -1 - \eta$. There are two kinds of traveling kinks:

$$\theta^> = 2\arctan[\pm\sqrt{\tfrac{4|\eta|-1}{4|\eta|+1}}\ \coth(\sqrt{\tfrac{16\eta^2 - 1}{16|\eta|}}\ s)] \tag{15}$$

$$\theta^< = 2\arctan[\pm\sqrt{\tfrac{4|\eta|-1}{4|\eta|+1}}\ \tanh(\sqrt{\tfrac{16\eta^2 - 1}{16|\eta|}}\ s)] \tag{16}$$

where the superscripts > and < denote the large kink and small kink respectively. We discuss their energies separately.

(A) Large kink: From (15) we found that $\theta \in [\phi_0, 2\pi - \phi_0]$, and it must vary cross one of the absolute maxima. In view of (8) and (11),

$$\begin{aligned}\langle H \rangle = 2V_s^> &= \sqrt{2}\int_{\phi_0}^{2\pi-\phi_0}\sqrt{-\eta - \tfrac{1}{8\eta} - \cos\theta - \eta\cos 2\theta}\ d\theta = \sqrt{2}\int_{\phi_0}^{2\pi-\phi_0}\sqrt{-2\eta(\cos\theta + \tfrac{1}{4\eta})^2}\ d\theta \\ &= \tfrac{1}{\sqrt{-\eta}}(\sqrt{16\eta^2 - 1} + \pi - \phi_0)\end{aligned} \tag{17}$$

(B) Small kink: From (16) we found that $\theta \in [-\phi_0, \phi_0]$, and it must contain one of the

relative maxima

$$\langle H \rangle = 2V_s^< = \sqrt{2}\int_{\phi_0}^{-\phi_0}\sqrt{-\eta - \frac{1}{8\eta} - \cos\theta - \eta\cos 2\theta}\,d\theta = \frac{1}{\sqrt{-\eta}}(\sqrt{16\eta^2 - 1} - \phi_0) \tag{18}$$

(p.s. there are small errors in (3.10) and (3.11) of [4].)

**Case 2**. $|\eta| < \frac{1}{4}$

There is only one type of baisc kink solution in this case:

$$\theta^> = 2\arctan[\pm\sqrt{1 + 4\eta}\,csch(\sqrt{1 + 4\eta}\,s] \tag{19}$$

The minimum of $V(\theta)$ are located at $\theta_{\min} = 2n\pi$ with $V(\theta_{\min}) = -1 - \eta$ and the maximum of $V(\theta)$ are located at $\theta_{\max} = (2n + 1)\pi$ with $V(\theta_{\max}) = 1 - \eta$. There is no relative maximum or minimum. We have $S = -V(\theta_{\min}) = 1 + \eta$ and from (8) and (11),

$$\langle H \rangle = 2V_s^> = \sqrt{2}\int_0^{2\pi}\sqrt{1 + \eta - \cos\theta - \eta\cos 2\theta}\,d\theta \tag{20}$$

The integral also depends on wether $\eta$ is larger or smaller than 0.

(A) $\frac{1}{4} > \eta > 0$

$$\langle H \rangle = 2V_s^> = 4\sqrt{4\eta + 1} + \frac{2\ln(2\sqrt{\eta} + \sqrt{4\eta + 1})}{\sqrt{\eta}} \tag{21}$$

(B) $0 > \eta > -\frac{1}{4}$

$$\langle H \rangle = 2V_s^> = 4\sqrt{4\eta + 1} + \frac{2\arcsin(2\sqrt{-\eta})}{\sqrt{-\eta}} \tag{22}$$

(p.s. there are small errors in (3.7) of [4].)

**Case 3** $\eta > \frac{1}{4}$

There are two kinds of traveling kinks:

$$\theta^> = 2\arctan[\pm\sqrt{1 + 4\eta}\,csch(\sqrt{1 + 4\eta}\,s] \tag{23}$$

and

$$\theta^B = 2\arctan[\pm\frac{1}{\sqrt{4\eta - 1}}\cosh(\sqrt{4\eta - 1}\,s)] \tag{24}$$

where the superscript B denotes the bubble solution. The absolute minimum of $V(\theta)$ are located at $\theta_{abs.\min} = 2n\pi$ with $V(\theta_{abs.\min}) = -1 - \eta$ and the maximum of $V(\theta)$ are located at $\theta_{\max} = \arccos(-1/4\eta) + 2n\pi$ with $V(\theta_{abs.\max}) = 1/8\eta + \eta$. The energy of the large kink is the same as that in case 2:

$$\langle H \rangle = 2V_s = 4\sqrt{4\eta + 1} + \frac{2\ln(2\sqrt{\eta} + \sqrt{4\eta + 1})}{\sqrt{\eta}} \tag{25}$$

The other kind of solution is the bubble solution. It extends from one relative minimum to another. These minima are at $\theta_{rel\min} \to (2n + 1)\pi$ as $s \to \pm\infty$. In this case, $S = -V(\theta_{rel,\min}) = \eta - 1$

$$\langle H \rangle = 2V_s^B = 2\sqrt{2}\int_{2\arctan(1/\sqrt{4\eta-1})}^{\pi}\sqrt{\eta - 1 - \cos\theta - \eta\cos 2\theta}\,d\theta = 4\sqrt{4\eta - 1} - \frac{4\ln(2\sqrt{\eta} + \sqrt{4\eta - 1})}{\sqrt{\eta}} \tag{26}$$

Note the upper bound and lower bound of the integral. We divide the bubble into two equal halves. The upper bound $\pi$ is the relative minimum while the lower bound is the middle point of the bubble. It comes from
$\theta^B(s=0) = 2\arctan[\pm(4\eta-1)^{-1/2}\cosh(\sqrt{4\eta-1}\,s)]|_{s=0} = \pm 2\arctan(1/\sqrt{4\eta-1})$. ( p.s. there are small errors in (3.7) and (3.9) of [4],too).

Further insight can be gained by applying (7). For example, if we start from $\varphi = 0$ and $\eta > \frac{1}{4}$, the solutions are the large kink in (23) and bubble in (24). When $\varphi$ is changed adiabatically into $\pi/2$, the solutions become the large and small kinks in (15) and (16):

$$\theta^> = 2\arctan[\pm\sqrt{\frac{4|\eta|-1}{4|\eta|+1}}\coth(\sqrt{\frac{16\eta^2-1}{16|\eta|}}\,s)] - \frac{\pi}{2} \tag{27}$$

$$\theta^< = 2\arctan[\pm\sqrt{\frac{4|\eta|-1}{4|\eta|+1}}\tanh(\sqrt{\frac{16\eta^2-1}{16|\eta|}}\,s)] - \frac{\pi}{2} \tag{28}$$

with the additional term $-\pi/2$ coming from the difference between $\theta$ and $\widetilde{\theta}$. Therefore, the solutions can have quite different forms as $\phi$ varies. The interesting question is whether the solution evolve smoothly or they change abruptly.

## 3. Analytical methods of asymmetry double SGE

### 3.1 The Möbius transformation from elliptic function ( genus=1) to the function of genus=0:

We propose that the solution of (1) in general has the form

$$\theta = 2\arctan[f(s)] \tag{29}$$

In Appendix, we have know the general transformation from one elliptic function ( genus=1) to another elliptic function ( genus=1). But look at Section 2, (19) and (23), in $|\eta|<\frac{1}{4}$ and $\eta > \frac{1}{4}$, the solution of DSGE withous phase is $\theta = 2\arctan[\pm\sqrt{1+4\eta}\,csch(\sqrt{1+4\eta}\,s]$. So it is reasonable to consider the following Möbius transformation to study the solution of the asymmetry DSGE:

$$f(s) = \frac{a\sinh(rs)+b}{c\sinh(rs)+d} \tag{30}$$

But both $\csc h$ and sinh function both are not elliptic function. So there will be some differences wirh Appendix. If we let $\varsigma = \sinh(rs)$, which satisfy $(\frac{d\varsigma}{ds})^2 = r^2(1+\varsigma^2) = r^2(\varsigma+i)\cdot(\varsigma-i)$. We do Möbius transformation let the roots of (A.1), $f_0, f_1, f_3$ correspond to $\varsigma_0 = i, \varsigma_1 = -i, \infty$. Then by Möbius transformation is

$$\frac{f_1-f_0}{f_1-f}\cdot\frac{f_3-f}{f_3-f_0} = \frac{\varsigma_1-\varsigma_0}{\varsigma_1-\varsigma} \tag{31}$$

So we have following transformation:

$$f(s) = \frac{f_3(f_1-f_0)\cdot\sinh(rs)+i[2f_0f_1-f_3(f_0+f_1)]}{(f_1-f_0)\cdot\sinh(rs)+i[f_0+f_1-2f_3]} \tag{32}$$

We require $\varsigma_2 = \varsigma_3 \to \infty$, so

$$\lambda = \frac{f_1-f_0}{f_1-f_2}\cdot\frac{f_3-f_2}{f_3-f_0} = \frac{\varsigma_1-\varsigma_0}{\varsigma_1-\varsigma_2}\cdot\frac{\varsigma_3-\varsigma_2}{\varsigma_3-\varsigma_0} \to 0 \tag{33}$$

This means the degenerate case, $f_2 = f_3$. This means in (30), $a = f_3(f_1-f_0)$ and $c = f_1-f_0$, etc.

### 3.2 Analytical methos of asymmetry DSGE:

After the substitution of (29) into (4), we have the following equation

$$(\frac{df}{ds})^2 = (\frac{S - \cos(\varphi) - \eta}{2})f^4 + \sin(\varphi)f^3 + (S + 3\eta)f^2 + \sin(\varphi)f + (\frac{S + \cos(\varphi) - \eta}{2}) \tag{34}$$

This is what Appendix concern because it is the differential equation of the elliptic function $f(s)$.
If we normalize (A.1) for $A = 1$, then $\varphi(s) = \prod_{i=0}^{3}(f - f_i) = f^4 - \theta_1 f^3 + \theta_2 f^2 - \theta_3 f + \theta_4$, and because $f_2 = f_3$, then

$$(1). \theta_1 = f_0 + f_1 + 2f_3 = -\frac{2\sin\varphi}{S - \cos\varphi - \eta}$$

$$(2). \theta_2 = f_0 f_1 + f_3^2 + 2f_3(f_0 + f_1) = \frac{2S + 6\eta}{S - \cos\varphi - \eta}$$

$$(3). \theta_3 = 2f_0 f_1 f_3 + f_3^2(f_0 + f_1) = -\frac{2\sin\varphi}{S - \cos\varphi - \eta}$$

$$(4). \theta_4 = f_0 f_1 f_3^2 = \frac{S + \cos\varphi - \eta}{S - \cos\varphi - \eta} \tag{35}$$

Let $t = f_0 + f_1$ and $s = f_0 \cdot f_1$, from (35)

$$t - \frac{2f_3}{1 - f_3^2} \cdot (s - 1) = 0 \tag{36a}$$

$$s - 1 + (f_3^2 - 1) \cdot (1 + \frac{\sin\varphi}{S - \cos\varphi - \eta} \cdot \frac{1}{f_3}) = 0 \tag{36b}$$

$$f_3^3 + \frac{3}{2} \cdot \frac{\sin\varphi}{S - \cos\varphi - \eta} \cdot f_3^2 + \frac{S + 3\eta}{S - \cos\varphi - \eta} \cdot f_3 + \frac{1}{2} \cdot \frac{\sin\varphi}{S - \cos\varphi - \eta} = 0 \tag{36c}$$

$$f_3^4 + \frac{\sin\varphi}{S - \cos\varphi - \eta} \cdot f_3^3 - \frac{\sin\varphi}{S - \cos\varphi - \eta} \cdot f_3 - \frac{S + \cos\varphi - \eta}{S - \cos\varphi - \eta} = 0 \tag{36d}$$

Finally, we want to solve the equations (36c) and (36d). $f_3$ and $S$ can be solved by analytical method because we can always find square solutions of polynomial of degree 4 in one variable. Following are the steps of "Euclidean algorithm".

### 3.3 Steps of "Euclidean algorithm" to solve (36c) and (36d):

(**1**). $2 \cdot (S - \cos\varphi - \eta) \cdot [f_3 \cdot (36c) - (36d)]$:

$$\sin\varphi \cdot f_3^3 + (2E + 6\eta) \cdot f_3^2 + 3\sin\varphi \cdot f_3 + 2(S + \cos\varphi - \eta) = 0 \tag{37}$$

(**2**). $(36c) - \frac{(37)}{\sin\varphi}$:

$$[2(2S + 6\eta)(S - \cos\varphi - \eta) - 3\sin^2\varphi] \cdot f_3^2 + 2\sin\varphi(2S - 3\cos\varphi - 6\eta) \cdot f_3 + [4(S + \cos\varphi - \eta)(S - \cos\varphi - \eta)$$

We simpily the symbols of (38) as

$$a_1 \cdot f_3^2 + b_1 \cdot f_3 + c_1 = 0 \tag{39}$$

(**3**). $\frac{a_1}{\sin\varphi} \cdot (37) - f_3 \cdot (38)$:

$$[b_1 - \frac{2S + 6\eta}{\sin\varphi} \cdot a_1] \cdot f_3^2 + [c_1 - 3\sin^2\varphi \cdot a_1] \cdot f_3 - 2(S + \cos\varphi - \eta) \cdot a_1 = 0 \tag{40}$$

We simpily the symbols of (40) as

$$a_2 \cdot f_3^2 + b_2 \cdot f_3 + c_2 = 0 \tag{41}$$

(**4**). $(39) \cdot a_2 - (41) \cdot a_1$:

$$A \cdot f_3 + B = 0 \tag{42}$$

where $A = b_1 \cdot a_2 - b_2 \cdot a_1$ and $B = c_1 \cdot a_2 - c_2 \cdot a_1$

(**5**). (39) $\cdot A -$ (42) $\cdot a_1 \cdot f_3$:

$$C \cdot f_3 + D = 0 \tag{43}$$

where $C = b_1 \cdot A - a_1 \cdot B$ and $D = c_1 \cdot A$

(**6**). **Both (42) and (43) will be satisfied**, then:

$$A \cdot D = B \cdot C \tag{44}$$

(44) is reduced to solve $S$. But in practical sense, it is too complexity. These six steps tell us if we solve $S$, then will get the degenerate roots:

$$f_2 = f_3 = -\frac{B}{A} = -\frac{D}{C} \tag{45}$$

**3.4 Another equivalent equation to solve $f_3$ and $S$:**

Compare (30) and (32), then $a = f_3(f_1 - f_0)$ and $c = f_1 - f_0$. At $s \to \pm\infty$, (29) give $\tan\frac{\theta}{2} = \frac{a}{c} = f_3$. Put into

$$-V|_{s\to\pm\infty} = [\cos(\theta + \varphi) + \eta\cos 2\theta]|_{\tan\frac{\theta}{2}=f_3} \tag{46}$$

We call $S = -V|_{s\to\pm\infty}$, then (46) give us

$$(\frac{S - \cos(\varphi) - \eta}{2})f_3^4 + \sin(\varphi)f_3^3 + (S + 3\eta)f_3^2 + \sin(\varphi)f_3 + (\frac{S + \cos(\varphi) - \eta}{2}) = 0 \tag{47}$$

But $f_3$ is the root of $\varphi(f)$ below (34). So (46) is consist with (35). And $S = -V|_{s\to\pm\infty} = -V_{\min}$ has been explained in (5) below. Using this switch point. It is also consist with to solve the extreme equation: $\frac{\partial V}{\partial \theta} = 0$

$$\frac{\partial V}{\partial \theta} = \sin(\theta + \varphi) + 2\eta\sin 2\theta = 0 \tag{48}$$

Also let $\tan\frac{\theta}{2} = f_3$ to solve (48) because it is the extreme case ( i.s $s \to \pm\infty$), then (48) become

$$f_3^4 + \frac{(8\eta - 2\cos\varphi)}{\sin\varphi} \cdot f_3^3 - \frac{(2\cos\varphi + 8\eta)}{\sin\varphi} \cdot f_3 - 1 = 0 \tag{49}$$

This pretty good equation is also consist with (35) and it can be solved by equation theory of one variable $f_3$. Here we will sketch the process how to solve (49).

**3.5 Steps to solve (49):**

In equation theory of degree 4 of one variable, we want to solve

$$g(f_3) = \prod_{j=1}^{4}(f_3 - \alpha_j) = f_3^4 - \Theta_1 \cdot f_3^3 + \Theta_2 \cdot f_3^2 - \Theta_3 \cdot f_3 + \Theta_4 \tag{50}$$

Compare with (49), then

$$\Theta_1 = -\frac{(8\eta - 2\cos\varphi)}{\sin\varphi}$$

$$\Theta_2 = 0$$

$$\Theta_3 = -\frac{(2\cos\varphi + 8\eta)}{\sin\varphi}$$

$$\Theta_4 = -1 \tag{51}$$

We can solve $\alpha_1 \sim \alpha_4$ by lower the degree as follows: Let $\beta_1 = \alpha_1\alpha_2 + \alpha_3\alpha_4$, $\beta_2 = \alpha_1\alpha_3 + \alpha_2\alpha_4$, $\beta_1 = \alpha_1\alpha_4 + \alpha_2\alpha_3$, let

$$h(x) = \prod_{j=1}^{3}(x - \beta_j) = x^3 - (\beta_1 + \beta_2 + \beta_3)x^2 + (\beta_1\beta_2 + \beta_1\beta_3 + \beta_2\beta_3)x - \beta_1\beta_2\beta_3$$

$$= x^3 - \Theta_2 \cdot x^2 + (\Theta_1 \cdot \Theta_3 - 4\Theta_4) \cdot x - \Theta_4 \cdot (\Theta_1^2 - 4\Theta_2) - \Theta_3^2$$

$$= x^3 + 4 \cdot \frac{1 - 16\eta^2}{\sin^2\varphi} \cdot x - 64 \cdot \frac{\eta\cos\varphi}{\sin^2\varphi} = 0 \quad (52)$$

Let $\Phi_2 = 4 \cdot \frac{1-16\eta^2}{\sin^2\varphi}$ and $\Phi_3 = 64 \cdot \frac{\eta\cos\varphi}{\sin^2\varphi}$, then (52) is $h(x) = x^3 + \Phi_2 \cdot x - \Phi_3$. The discriminant $\Delta$ is defined as

$$\Delta = \frac{\Phi_3^2}{4} + \frac{\Phi_2^3}{27} = 64[-\frac{16\eta^2}{\sin^2\varphi} + \frac{16\eta^2}{\sin^4\varphi} + \frac{1}{27} \cdot \frac{(1-16\eta^2)^3}{\sin^6\varphi}] \quad (53)$$

The three roots of $h(x) = 0$ are

$$\beta_1 = \sqrt[3]{\frac{4\eta + \cos\varphi}{\sin\varphi} + \sqrt{\Delta}} + \sqrt[3]{\frac{4\eta + \cos\varphi}{\sin\varphi} - \sqrt{\Delta}}$$

$$\beta_2 = \sqrt[3]{\frac{4\eta + \cos\varphi}{\sin\varphi} + \sqrt{\Delta}} \cdot \zeta^2 + \sqrt[3]{\frac{4\eta + \cos\varphi}{\sin\varphi} - \sqrt{\Delta}} \cdot \zeta$$

$$\beta_3 = \sqrt[3]{\frac{4\eta + \cos\varphi}{\sin\varphi} + \sqrt{\Delta}} \cdot \zeta + \sqrt[3]{\frac{4\eta + \cos\varphi}{\sin\varphi} - \sqrt{\Delta}} \cdot \zeta^2 \quad (54)$$

Where $\zeta^3 = 1, \zeta \neq 1$. Define:

$$\gamma_1^2 = \Theta_1^2 - 4 \cdot \Theta_2 + 4\beta_1 = 4(\frac{4\eta - \cos\varphi}{\sin\varphi})^2 + 4\beta_1$$

$$\gamma_2^2 = \Theta_1^2 - 4 \cdot \Theta_2 + 4\beta_2 = 4(\frac{4\eta - \cos\varphi}{\sin\varphi})^2 + 4\beta_2$$

$$\gamma_3^2 = \Theta_1^2 - 4 \cdot \Theta_2 + 4\beta_3 = 4(\frac{4\eta - \cos\varphi}{\sin\varphi})^2 + 4\beta_3 \quad (55)$$

And choose the sign of $\gamma_1, \gamma_2, \gamma_3$ to satisfy
$\gamma_1 \cdot \gamma_2 \cdot \gamma_3 = 8\Theta_3 - 4\Theta_1 \cdot \Theta_2 + \Theta_1^3 = 16(\frac{4\eta+\cos\varphi}{\sin\varphi}) - 8(\frac{4\eta-\cos\varphi}{\sin\varphi})^3$. So the four extreme points of (50) are:

$$\alpha_1 = \frac{1}{4}(\Theta_1 + \gamma_1 + \gamma_2 + \gamma_3)$$

$$\alpha_2 = \frac{1}{4}(\Theta_1 + \gamma_1 - \gamma_2 - \gamma_3)$$

$$\alpha_3 = \frac{1}{4}(\Theta_1 - \gamma_1 + \gamma_2 - \gamma_3)$$

$$\alpha_4 = \frac{1}{4}(\Theta_1 - \gamma_1 - \gamma_2 + \gamma_3) \quad (56)$$

One of (56) are the analytical solutions of $f_3 = -V|_{\min} = -V|_{s \to \pm\infty}$. Each one of (56) are the possible solution of $f_3$ because with different $\eta$ and $\varphi$. The absolute minimum with switch from $\alpha_i$ to $\alpha_j$. The best way to decide which one is $f_3$ is to take each $\alpha_i$ of (56) into $-V = \cos(\theta + \varphi) + \eta\cos 2\theta$ and see which one is the minimum. And this value is $S = -V|_{\min}$. We also remind the reader that once we get the analytical solution of $f_3$ and $S$, we can put into (35) to get the analytical expression of $f_0$ and $f_1$, too. And go back to (32) to get the analytical expression of $a, b, c, d$. Which means we get the analytical expression of asymmetry DSGE by Möbius transformation, (30).

**4. Combine equations of to decide** $a, b, c, d, S, r$ **together**:

In [20], We use numerical method to decide $a, b, c, d, S, r$ together with little correction. Substituting (29) and (30) into (4) and requiring the scaling

$$(ad - bc)^2 = 1 \tag{57a}$$

we have found the following equations by comparing the powers of sinh function

$$(a^2 + c^2)^2 S + (c^4 - a^4) \cos \varphi + 2ac(a^2 + c^2) \sin \varphi + (-a^4 + 6a^2c^2 - c^4)\eta = 0 \tag{57b}$$

$$\begin{aligned} & 4(ab + cd)(a^2 + c^2)S - 4(a^3b - c^3d) \cos \varphi \\ & + 2(a^3d + 3a^2bc + 3ac^2d + bc^3) \sin \varphi \\ & + [-4a^3b + 6(2a^2cd + 2abc^2) - 4c^3d]\eta = 0 \end{aligned} \tag{57c}$$

$$\begin{aligned} & [6a^2b^2 + 2(a^2d^2 + 4abcd + b^2c^2) + 6c^2d^2]S \\ & - (6a^2b^2 - 6c^2d^2) \cos \varphi \\ & + 6(a^2bd + ab^2c + acd^2 + bc^2d) \sin \varphi \\ & + 6(-a^2b^2 + a^2d^2 + 4abcd + b^2c^2 - c^2d^2)\eta = 2r^2 \end{aligned} \tag{57d}$$

$$\begin{aligned} & 4(ab + cd)(b^2 + d^2)S - 4(b^3a - d^3c) \cos \varphi + \\ & 2(b^3c + 3b^2ad + 3bd^2c + ad^3) \sin \varphi + \\ & [-4b^3a + 6(2d^2ab + 2cdb^2) - 4d^3c]\eta = 0 \end{aligned} \tag{57e}$$

$$(b^2 + d^2)^2 S + (d^4 - b^4) \cos \varphi + 2bd(b^2 + d^2) \sin \varphi + (-b^4 + 6b^2d^2 - d^4)\eta = 2r^2 \tag{57f}$$

The simultanous algebrac equations can be solved to give $a$, $b$, $c$, $d$, $r$ and $S$ together. (57) are clear and short, but maybe we can tangle them by numerical method not analytical method like Sec. 3.

The special cases are generic and very intersting. It happens at $\varphi = \pi/2$ and $\eta > 1/4$. We explain their cause by starting with the symmetry of the DSGE with additional phase. If one substitutes $\pi - \phi$ for $\phi$, then he can replace $\theta$ with $\pi - \theta$ and (1) retains its original form. In view of (57), we have the following symmetry:

$$\begin{aligned} \phi & \rightarrow \pi - \phi \\ a & \rightarrow -c \\ b & \rightarrow d \\ c & \rightarrow -a \\ d & \rightarrow b \\ r & \rightarrow r \\ S & \rightarrow S \end{aligned} \tag{58}$$

Suppose we have, and indeed we have found, real solution of (57) for $\eta > \frac{1}{4}$ with $a$, $b$, $c$ and $d$ being continuous with varying $\phi$, then we must have $a = -c$ and $b = d$ at $\varphi = \pi/2$. Substituting these into (57), we get

$$a^4(S - 1 + \eta) = 0 \tag{59a}$$

$$a^2 b^2 = \frac{r^2}{4S - 12\eta} \tag{59b}$$

$$b^4(S + 1 + \eta) = \frac{r^2}{2} \tag{59c}$$

along with the scaling restriction from (57a)

$$4a^2b^2 = 1 \tag{59d}$$

Eq. (59d) means $a \neq 0$, and so (59a) implies $S + \eta = 1$. Inserting these into (59b) and (59c), we get

$$\frac{r^2}{1 - 4\eta} = 1 \tag{59e}$$

We get contradict if $\eta > \frac{1}{4}$.

## 5. Discussion and conclusion

In this work, we used the analytical method of "Möbius transformation" to solve the DSGE with an phase $\phi$. And this paper is go further beyond [20], where in their we use the numerical method.

## Appendix: Introduction to Möbius transformation

By Jacobi elliptic function theory, one can transform the following equation:

$$(\frac{df}{ds})^2 = \varphi(s) = A(f - f_0)(f - f_1)(f - f_2)(f - f_3), \tag{A-1}$$

where $\varphi(s)$ is a polynomial of $s$ to the three or four power and $f_0, f_1, f_2$ and $f_3$ are the roots, into the standard form, i. e., only terms with even powers.present. The "Möbius transformation" has the form

$$f = \frac{a\zeta + b}{c\zeta + d}, \tag{A-2}$$

and so does every root

$$f_i = \frac{a\zeta_i + b}{c\zeta_i + d}. \tag{A-3}$$

If we take special values of $a$, $b$, $c$ and $d$, we can obtain the form

$$f = \frac{f_3(f_1 - f_0) * \zeta - f_1(f_3 - f_0)}{(f_1 - f_0)\zeta - (f_3 - f_0)} \tag{A-4}$$

and eq. (A-1) becomes

$$(\frac{d\varsigma}{ds})^2 = B(\varsigma - \beta_0)(\varsigma - \beta_1)(\varsigma - \beta_2)(\varsigma - \beta_3). \tag{A-5}$$

We set

$$\lambda = \frac{f_1 - f_0}{f_1 - f_2}\frac{f_3 - f_2}{f_3 - f_0} = \frac{\beta_1 - \beta_0}{\beta_1 - \beta_2}\frac{\beta_3 - \beta_2}{\beta_3 - \beta_0}, \tag{A-6}$$

so that eq. (A-5) becomes

$$(\frac{d\varsigma}{ds})^2 = B\varsigma(\varsigma - 1)(\lambda\varsigma - 1) \tag{A-7}$$

here $B = A(f_3 - f_0)(f_2 - f_1)$. This is the standard form. But we can do further transform by setting $\varsigma = \xi^2$. So eq. (A-7) becomes the differential equation of Jacobi elliptic function:

$$(\frac{d\xi}{ds})^2 = \frac{B}{4}(\xi^2 - 1)(\lambda\xi^2 - 1). \qquad \text{A-8}$$

Instead eqs. (A.4) and (A.8), we can take different transformation by mapping roots into $1, -1, 1/k,$ and $-1/k$. So

$$\lambda = \frac{f_1 - f_0}{f_1 - f_2} \frac{f_3 - f_2}{f_3 - f_0} = \frac{\frac{1}{k} - 1}{\frac{1}{k} + 1} \frac{\frac{-1}{k} + 1}{\frac{-1}{k} - 1} = (\frac{1-k}{1+k})^2 \qquad \text{A-9}$$

and eq. (A-1) becoms

$$(\frac{d\xi}{ds})^2 = B'(\xi^2 - 1)(k^2\xi^2 - 1) \qquad \text{A-10}$$

There are more details to be solved such as how to deal with degenerate roots and how to further transform $\lambda$ so that it is real and in the range $(0, 1)$ in order that eq. (A-8) and (A-10) are compatible with the Jacobi elliptic differential equation. But these are beyond the scope of this work.